\documentclass[apjl]{emulateapj}
\usepackage{apjfonts}

\shorttitle{Galaxy Properties vs. Host Halo Mass}
\shortauthors{Shankar et al.} 

\begin{document}

\title{New Relationships between Galaxy Properties and Host Halo
Mass,\\ and the Role of Feedbacks in Galaxy Formation}

\author{F. Shankar\altaffilmark{1}, A. Lapi\altaffilmark{1}, P. Salucci\altaffilmark{1},
G. De Zotti\altaffilmark{2,1} \& L. Danese\altaffilmark{1}}
\altaffiltext{1}{Astrophysics Sector, SISSA/ISAS, Via Beirut 2-4,
I-34014 Trieste, Italy}\altaffiltext{2}{INAF, Osservatorio
Astronomico di Padova, Vicolo dell'Osservatorio 5, I-35122 Padova,
Italy}

\begin{abstract}
We present new relationships between halo masses ($M_h$) and several
galaxy properties, including $r^*$-band luminosities ($L_r$),
stellar ($M_{\mathrm{star}}$) and baryonic masses, stellar velocity
dispersions ($\sigma$), and black hole masses ($M_{\rm BH}$).
Approximate analytic expressions are given. In the galaxy halo mass
range $3\times 10^{10}\, M_{\odot}\leq M_h\leq 3\times 10^{13}\,
M_{\odot}$ the $M_h$--$L_r$, $M_{\mathrm{star}}$--$M_h$, and
$M_{\mathrm{BH}}$--$M_h$ are well represented by a double power law,
with a break at $M_{h,{\mathrm{break}}}\approx 3 \times 10^{11}\,
M_{\odot}$, corresponding to a mass in stars $M_{\mathrm{star}}\sim
1.2\times 10^{10}\, M_{\odot}$, to a $r^*$-band luminosity $L_r\sim
5 \times 10^9\, L_{\odot}$, to a stellar velocity dispersion $\sigma
\simeq 88$ km s$^{-1}$, and to a black hole mass $M_{BH}\sim 9\times
10^{6}\,M_{\odot}$. The $\sigma$--$M_h$ relation can be approximated
by a single power law, though a double power law is a better
representation. 
Although there are
significant systematic errors associated to our method, the derived
relationships are in good agreement with the available observational
data and have comparable uncertainties. We interpret these relations
in terms of the effect of feedback from supernovae and from the
active nucleus on the interstellar medium. We argue that the break
of the power laws occurs at a mass which marks the transition
between the dominance of the stellar and the AGN feedback.

\end{abstract}


\section{Introduction}

There is a well determined set of cosmological parameters $h\equiv
H_0/100 \, \mathrm{km\, s}^{-1} \, \mathrm{Mpc}^{-1}=0.70 \pm
0.04$, $\Omega_M = 0.27 \pm 0.04$, $\Omega_{\Lambda} = 0.73 \pm
0.05$, $t_0=13.7 \pm 0.2$ Gyr, $\sigma_8=0.84\pm 0.04$ emerging
from a number of observations (the concordance cosmology, see
Spergel et al. 2003). Also the cosmic density of baryons
$\Omega_{b} = 0.044 \pm 0.004$ has been very precisely determined
through both the power spectrum of Cosmic Microwave Background
anisotropies and measurements of the primordial abundance of light
elements (Cyburt et al. 2001; Olive 2002). An important
complementary information is that the density of baryons residing
in virialized structures and associated to detectable emissions is
much smaller than $\Omega_{b}$. In fact, traced-by-light baryons
in stars and in cold gaseous disks of galaxies and in hot gas in
clusters amount to a $\Omega_{b,\mathrm{lum}} \approx (3-4) \times
10^{-3}\la 0.1 \Omega_{b}$ (Persic \& Salucci 1992, Fugukita et
al. 1998, Fukugita \& Peebles 2004). On the other hand, in rich
galaxy clusters the ratio between the mass of the dark matter (DM)
component and the mass of the baryon component, mainly in the hot
intergalactic gas, practically matches the cosmic ratio
$\Omega_M/\Omega_b$.

The circumstance that $\Omega_{b}$ is a factor of about 10 larger
than $\Omega_{b,\mathrm{lum}}$ puts forth both an observational
and a theoretical problem. On one side, observations are needed to
detect and locate these ``missing'' baryons (see for a review
Stocke, Penton \& Shull 2003). On the theoretical side, galaxy
formation models have to cope with the small amount of baryons
presently in gas and stars inside galaxies. No doubt that feedback
from stars and AGNs played a relevant role in unbinding large
amounts of gas and eventually removing them from the host DM halo
(see, e.g., Dekel \& Silk 1986; Silk \& Rees 1998; Granato et al.
2001, 2004; Hopkins et al. 2005; Lapi, Cavaliere \& Menci 2005),
but we need to get a detailed quantitative understanding of these
processes, which are crucial to comprehend galaxy formation. The
stellar feedback is expected to depend on the star formation
history and the total energy released to the gas ultimately
depends on the total mass of formed stars and on the present-day
galaxy luminosity. Correspondingly, the total energy injected by
the AGN feedback is ultimately related to the final black hole
(BH) mass. The fraction of the gas removed by feedbacks is
expected to depend also on the binding energy of the gas itself,
which is determined by the galaxy \emph{virial} mass and by its
density distribution. Therefore the relationships between the
galaxy halo mass and the galaxy luminosity, the stellar and
baryonic mass, the mass of the central BH, are expected to give
extremely useful information on the process of galaxy formation
and evolution. An additional relevant piece of information is the
link between the galaxy halo mass and the velocity dispersion of
the old stellar component. This paper is devoted to a statistical
study of these relations.

The Halo Occupation Distribution (see, e.g., Kauffmann, Nusser \&
Steinmetz 1997; Peacock \& Smith 2000; Berlind \& Weinberg 2002;
Magliocchetti \& Porciani 2003), which specifies the probability
$P(N|M)$ that a halo of mass $M$ is hosting $N$ galaxies, is a
helpful statistics to establish the link between the host DM halo
mass and the observed galaxy luminosity. An additional tool to
explore this link is the formalism of the Conditional Luminosity
Function (Yang, Mo \& van den Bosch 2003), which describes how
many galaxies of given luminosity reside in a halo of given mass.
Following this approach, Yang et al. (2003) investigated the
relation between halo mass and luminosity. However, particularly
for high halo masses, $M_h\ga 10^{13}\, M_{\odot}$, both methods
give information on galaxy systems, more than on large galaxies.

Marinoni \& Hudson (2002) and Vale \& Ostriker (2004) suggest that
a helpful starting point can be the simple hypothesis that there
is a one-to-one, monotonic correspondence between halo mass and
resident galaxy luminosity. Then, by equating the integral number
density of galaxies as a function of their luminosity and stellar
mass to the number density of galaxy halos, one gets a statistical
estimate of the DM halo mass associated to galaxies of fixed
luminosity or fixed baryon/stellar mass. However, a major problem
of the method is the estimate of the mass function of halos
hosting one single galaxy, the Galaxy Halo Mass Function (GHMF).
To solve the problem, in this paper we use an empirical approach,
which takes into account the results of numerical simulations
(see, e.g., De Lucia et al. 2004 and references therein) on the
halo occupation distribution by adding to the Halo Mass Function
(HMF) the contribution of subhalos (Vale \& Ostriker 2004; van den
Bosch, Tormen \& Giocoli 2005). At large masses we subtract from
the HMF the mass function of galaxy groups (Girardi \& Giuricin
2000; Martinez et al. 2002; Hein\"am\"aki et al. 2003; Pisani,
Ramella \& Geller 2003).

The mass around a galaxy up to a radial distance from its center
much larger than the characteristic scale of light distribution
can be inferred from detailed X-ray observations (see, e.g.,
O'Sullivan \& Ponman 2004). Also  the statistical  measurements of
the shear induced by weak gravitational lensing around galaxies
(see Bartelmann, King, \& Schneider 2001) yield important insights
on the halo mass of galaxies (McKay et al. 2002, Sheldon et al.
2004). Though these mass estimates  have significant uncertainties
and their extrapolations to the virial radii are not immediate,
they nonetheless provide useful reference values to which we
compare the outcomes of our method.

The role of stellar and AGN feedback has been discussed by several
authors (see, e.g., Dekel \& Silk 1986; Silk \& Rees 1998).
Granato et al. (2001, 2004) have implemented both feedbacks  in
their model of joint formation of QSOs and spheroidal galaxies.
More recently the feedbacks have also been introduced into
hydrodynamical simulations (Springel, Di Matteo \& Hernquist
2005). One of the purposes of this paper is  to show how the
information coming from the relationships of the galaxy halo mass
with measurable galaxy properties (such as the stellar, baryonic
and central BH masses) can shed light on the role and on relative
importance of the two feedbacks.

The plan of this work is the following. In \S$\,2$ we compute the
galaxy stellar and baryonic mass functions, exploiting the
luminosity function and the mass-to-light ratio of the stellar
component inferred from kinematical mass modelling of galaxies.
Then, in \S$\,3$, we derive the mass function of galactic halos,
and, exploiting the relevant galaxy statistics (luminosity
function, galaxy star/baryon mass function, velocity dispersion
function) and the galaxy halo mass function, we investigate the
relationships of the corresponding galaxy properties with the halo
mass. The relation of the halo mass with the mass of the central
BH in galactic spheroidal components is deduced in \S$\,4$, by
comparing the central black hole mass function to the galaxy halo
mass function. In \S$\,5$ we discuss the role of the stellar and
AGN feedbacks in shaping the relationships between stellar and
baryonic mass and halo mass. \S$\,6$ is devoted to the
conclusions.

\section{The star and baryon mass functions of galaxies}

The luminosity function (LF) is a fundamental statistics for
galaxies. Its present form is the result of physical processes
involving both baryons and dark matter. In particular, the LFs in
the range between about 0.1 to several $\mu$m probe the stellar
component. The mass of stars and baryons associated to galaxies
can be derived coupling the LF with estimates of the mass-to-light
ratio (MLR) of the stellar and gaseous components. As it is well
known, the MLR and the fraction of gas depend on galaxy
morphology.

Nakamura et al. (2003) estimated the LF in the $r^*$-band for
early- and late-type galaxies separately. The separation has been
done through a light concentration method (see Shankar et al. 2004
for a discussion and a comparison with other LF estimates). These
early- and late-type galaxy LFs are in reasonable agreement with
the LFs of red and blue galaxies, respectively, derived by Baldry
et al. (2004).

Since the Nakamura LF of late-type galaxies is well defined only
for luminosities brighter than $M_r\la -18$, we extended it to
lower luminosity using the results of Zucca et al. (1997) and
Loveday (1998) and translating them from the $b_{J}$-band to the
$r^*$-band using a color $(b_J-r^*)\approx 0.33$, as appropriate
for star forming irregular galaxies (Fukugita et al. 1995; their
Table 3, panels (j) and (m) with $b_J\equiv B_J$ and $r^*\simeq
r'$). The conversion to solar luminosities has been done taking
$M_{r\odot}=4.62$ (Blanton et al. 2001). The resulting LF is well
fit, in the range $3\times 10^7\,L_{\odot}\la L_r\la 3\times
10^{11}\,L_{\odot}$, by
\begin{equation}\label{eq|LF}
{\phi(L_{r})\, \mathrm{d} L_{r}\over \hbox{Mpc}^{-3}} =
\left({9.05\times 10^{-3}\over x^{1.14}\, e^{0.0076\,x}}+ {4\times
10^{-5} \over x^{4.03}}\right) \mathrm{d}x~,
\end{equation}
where $x\equiv L_r/2.4\times 10^8\,L_{\odot}$.

The MLR of the stellar component can be derived from studies of
stellar evolution, with uncertainties associated to the poor
knowledge of details of the IMF (see, e.g., Fukugita et 1998; Bell
et al. 2003; Fukugita \& Peebles 2004; Baldry et al. 2004; Panter,
Heavens \& Jimenez 2004). A more direct approach exploits detailed
kinematical and photometric studies of galaxies to estimate the
amount of mass traced by light and the mass of the DM component,
taking advantage of their different distribution inside the
galaxies. This method has been used by Salucci \& Persic (1999), who
estimated the stellar and gaseous mass as a function of the B-band
luminosity for late-type galaxies to get the baryon mass $M_b\approx
1.33 M_{\rm HI}+M_{\mathrm{star}}$. We have approximated their
results for the stellar and the gas component, respectively as
\begin{equation}\label{eq|MsLB}
\log{{M_{\mathrm{star}}\over M_{\odot}}}=-1.6+1.2\,\log{{L_B\over
L_{\odot}}}~,
\end{equation}
for $10^7 \, M_{\odot}\leq M_{\rm star}\leq 10^{12} \, M_{\odot}$,
and
\begin{equation}\label{eq|MgLB}
\log{{M_{\rm HI}\over M_{\odot}}}=1.34 +0.81 \log{{L_B\over
L_{\odot}}}~
\end{equation}
in the range $3\times 10^6 \, M_{\odot}\leq M_{\rm HI}\leq 10^{11}
\, M_{\odot}$. Combining these relations with the LF of
eq.~(\ref{eq|LF}), assuming a Gaussian distribution around the mean
relations with a dispersion of about 20\%, we derived the stellar
mass function and the baryonic mass function of late-type galaxies.
To do that we have taken $M_{B\odot}=5.48$ (Binney \& Merrifield
1998) and $(B-r')=0.9$ (see Table 3, panel (m) of Fukugita et al.
1995).

A similar approach can be followed for early-type galaxies. For
about 9000 such galaxies extracted from the Sloan Digital Sky Survey
(SDSS) Bernardi et al. (2003) estimated the MLR (within the
characteristic half luminosity ratio $r_e$, in solar units) $M/L=3.6
\langle L/L_{\star} \rangle ^{0.15}$ $M_{\odot}/L_{\odot}$ in the
$r^*$-band ($L_{\star}=2\times 10^{10}\, L_{\odot}$). These authors
derived the mass inside $r_e$ using the relation $M(r_e)=
c\,\sigma^2\, r_e/G$, where $\sigma$ is the central velocity
dispersion, and assuming $c=2$. However, the value of $c$ depends on
the light profile ($c\approx 2.35$ for a de Vaucouleurs profile, see
Prugniel and Simien 1997) and on the DM distribution (Borriello et
al. 2003). Using $c=2.35$ and rescaling the zero point of the MLR by
Bernardi et al. (2003), we obtain:
\begin{equation}\label{eq|MsLr}
\left\langle\frac{M}{L}\right\rangle_{\mathrm{star}}=4.1\,\left
\langle{L\over L_{\star}}\right \rangle^{0.15}\,
\frac{M_{\odot}}{L_{\odot}}~.
\end{equation}
As discussed by Bell et al. (2003) and by Baldry et al. (2004),
the uncertainties related to the IMF and to the star formation
history imply an uncertainty of about $30\%$ in the mean value of
the MLR; the dispersion around the mean relation,
eq.~(\ref{eq|MsLr}), is of about $\sim 20\%$. By convolving the
$r^*$-band luminosity function of early-type galaxies of Nakamura
et al. (2003) with the distribution of MLRs, assuming a Gaussian
distribution around the mean relation of eq.~(\ref{eq|MsLr}), with
a dispersion of 20\%, we estimate the Galaxy Stellar Mass Function
(GSMF) of $E/S0$ galaxies, which essentially coincides with the
Galaxy Baryonic Mass Function (GBMF), since in early-type galaxies
the gas gives a negligible contribution to the baryon mass.

\begin{figure}[t]
\plotone{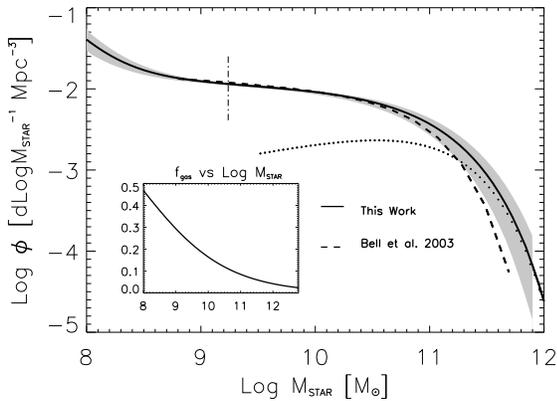}\caption{The Galaxy Stellar Mass Function. The thick
\textit{solid} line shows the numerical results, which are almost
perfectly matched by the analytic fitting formula
[eq.~(\protect\ref{eq|GSMF})]. The \textit{dotted} line gives the
contribution of early-type galaxies. The \textit{shaded} area
represents the uncertainty due to the $\simeq 30\%$ error in the
mean stellar mass to light ratio. The vertical \textit{dot-dashed}
line shows the stellar mass corresponding to $M_r=-18$. The
\textit{dashed} line shows the result by Bell et al. (2003). The
inset illustrates the average fraction of gas as a function of the
stellar mass.}\label{fig|GSMF}
\end{figure}

The total GSMF, holding over the mass range $10^{8}\, M_{\odot} \la
M_{\mathrm{star}} \la 10^{12}\, M_{\odot}$, is shown by the solid
line in Fig.~\ref{fig|GSMF}, where the shaded area corresponds to
the 30\% uncertainty in the mean MLR. This is a safe range to
determine the GSMF as at lower masses the uncertainty in the LF
grows while the increase in the total stellar mass density including
$M_{\rm star}<10^8\, M_{\odot}$ is rather small, $< 5\%$. The upturn
at $M_{\mathrm{star}}\la 3\times 10^8\, M_{\odot}$ corresponds to
the appearance of the dwarf galaxy population, represented by the
second term at r.h.s. of eq.~(\ref{eq|LF}). However the contribution
in stellar mass density in the range $10^{8}\, M_{\odot} \la
M_{\mathrm{star}} \la 10^{9}\, M_{\odot}$ is just $\sim$ 3\% of the
total. In the inset of Fig.~\ref{fig|GSMF} we have displayed the gas
fraction as a function of the stellar mass. A very accurate
analytical representation (actually indistinguishable from the solid
line showing the numerical results) is provided by a Schechter
function plus a power law term:
\begin{equation}\label{eq|GSMF}
{\mathrm{GSMF} (M_{\mathrm{star}})\, \mathrm{d}
M_{\mathrm{star}}\over \hbox{Mpc}^{-3}} = \left({3\times
10^{-3}\over x^{1.16}\, e^{0.32\,x}}+ {2.25\times 10^{-9} \over x^{
3.41}}\right) \mathrm{d}x~\, ,
\end{equation}
where $x\equiv M_{\mathrm{star}}/6\times 10^{10}\,M_{\odot}$. The
GBMF is easily computed by adding the appropriate gas contribution.
Recent estimates of the GSMF and of the GBMF have been produced by
Cole et al. (2001), Bell et al. (2003) and Baldry et al. (2004),
exploiting 2dF, SDSS and Two Micron All Sky Survey data. The
estimate by Bell et al. (2003) is shown by the dashed line in
Fig.~\ref{fig|GSMF}. The difference with our estimate is mainly due
to the difference in adopted MLR. Bell et al. (2003) have derived
their MLR by fitting the broad band SED with stellar population
models. The mean MLR adopted here, based on kinematic
determinations, is a factor of about $1.3$ higher at high
luminosities, while at very low luminosities ($L_r\la 5\times 10^8\,
L_{\odot}$) it is about a factor of $2$ lower. However, the flatness
of the GSMF at small masses conceals the difference in the MLR,
while at large masses the almost exponential decline of the LF
amplifies it. It is worth noticing that the determination of the MLR
of low luminosity objects is hampered by many effects related to the
episodic star formation history, to the presence of dust, to the
irregularity of their shapes, and to the DM predominance.

Cole et al. (2001) presented estimates of the GSMF for two choices
of the Initial Mass Function (IMF): Salpeter's (1955) and
Kennicutt's (1983). For a Salpeter IMF their GSMF is consistent
with ours, within our estimated uncertainties: the main difference
is a $\simeq 30\%$ excess for $\log(M_{\rm star}/M_\odot) \la
10.5$. For a Kennicutt IMF, their GSMF drops for $M_{\rm star}$
about 0.2 dex lower than that by Bell et al. (2003). The estimate
by Baldry et al. (2004) is close to that  by Bell et al. (2003),
as expected since they exploit very similar LFs and MLRs. All in
all, methods based on kinematic measurements and on stellar
population synthesis yield GSMFs and GBMFs in reasonable
agreement, and establish a sound confidence interval.

Integrating the GSMF for $M_{\mathrm{star}} \ga 10^8\, M_{\odot}$,
the mass density parameter of baryons condensed in stars
associated to late-type galaxies is found to be
\begin{equation}\label{eq|omegalate}
\Omega_{\mathrm{star}}^{L}(\mathrm{kin})=(1.8 \pm 0.4) \times
10^{-3}~,
\end{equation}
where the label ``kin'' indicates that the stellar mass of
galaxies has been estimated using kinematic data. The
corresponding neutral gas density amounts to $\sim 20\%$ of this
value and it is concentrated in late-type, low mass systems with
$M_{\mathrm{star}} \la 5\times 10^9\, M_{\odot}$. Here and in the
following eqs.~(\ref{eq|omegaearly}) and (\ref{eq|omegagal}), the
errors reflect the uncertainties on the GSMF.

The star density parameter associated to early-type galaxies amounts
to
\begin{equation}\label{eq|omegaearly}
\Omega_{\mathrm{star}}^{E}(\mathrm{kin})= (1.8 \pm 0.6) \times
10^{-3}~.
\end{equation}
It is well known that in early-types the amount of cold gas is
negligible. Therefore, the overall local stellar mass density in
galaxies with stellar masses in the range $10^8\, M_\odot \la M_b
\la 10^{12}\, M_\odot$ is
\begin{equation}\label{eq|omegagal}
\Omega_{\mathrm{star}}^{G} (\mathrm{kin})=(3.6 \pm 0.7) \times
10^{-3}~.
\end{equation}
This value is in good agreement with the recent estimates obtained
through spectro-photometric galaxy models (Bell et al. 2003;
Fukugita 2004; Fukugita \& Peebles 2004). The contribution of the
cold gas to the baryon density in galaxies is only $\sim 8\%$, and
thus $\Omega_{b}^{G}\approx 1.08 \Omega_{\mathrm{star}}^{G}$. This
result confirms the well known conclusion that only a small
fraction, $\la 10 \%$, of the cosmic baryons is today in stars and
cold gas within galaxies. It is worth noticing that the star
formation rate integrated over the cosmic history matches the
overall local mass density in stars (see Nagamine et al. 2004).
This mass density has been accumulated at high redshifts, $z\ga
1$, for early-type galaxies and later on for late-types, as
indicated by their respective stellar populations.

By subtracting from the cosmic matter density the contribution of
baryons ($\Omega_b \leq 0.044$) and that of dark matter (DM) in
groups and clusters of galaxies ($\Omega_{DM}^{\mathrm{C}}\approx
0.012$; Reiprich \& Bohringer 2002), we obtain the DM mass density
associated with galaxies $\Omega_{DM}^{G}\approx 0.21$, in
excellent agreement with the determination by Fukugita \& Peebles
(2004). The average DM-to-baryon (essentially stars) mass fraction
in galaxies turns out to be around $60$. This value must be
compared with the cosmological ratio
$R_{\mathrm{cosm}}=\Omega_{DM}/\Omega_{b}\approx 6$. In fact, in
rich galaxy clusters the baryon mass, mostly in form of diffuse
gas, and the DM halo mass have a ratio consistent with the
``cosmic'' DM to baryon ratio (see, e.g., Ettori, Tozzi \& Rosati
2003). This evidences that the baryon fraction in galaxies
decreases on average by a factor of about $10$ relative to the
initial value, due to a number of astrophysical processes
associated to the formation of these objects. In the following we
will use the cosmic fraction
$f_{\mathrm{cosm}}=1/R_{\mathrm{cosm}}\approx 0.17$.

\section{The Galaxy Halo Mass Function and the $L$, $M_{\mathrm{star}}$
and $\sigma$ vs. halo mass relations} \label{sect:ghmf}

In order to investigate the relationships between the stellar and
baryonic mass and the halo mass in galaxies in a one-to-one
correspondence, the statistics of halos containing {\it one} single
galaxy, the Galaxy Halo Mass Function (GHMF), has to be estimated.
The overall HMF as found by numerical simulations (see, e.g.,
Jenkins et al. 2001; Springel et al. 2005) is well described by the
Press \& Schechter (1974) formula as modified by Sheth \& Tormen
(1999). However, in order to compute the GHMF, we have to deal with
the problem of the halo occupation distribution (HOD; Peacock \&
Smith 2000; Berlind et al. 2003; Magliocchetti \& Porciani 2003;
Kravtsov et al. 2004; Abazajian et al. 2005). Two effects need to be
taken into account: (i) the number of galaxies is actually larger
than the number of DM halos, since a halo may contain a number of
sub-halos, each hosting a galaxy, and (ii) the probability that very
massive halos ($M_h \ga 10^{12}$--$10^{13}\, M_{\odot}$) host a
single giant galaxy drops rapidly with increasing halo mass.

To account for the effect (i) we use the results by Vale \& Ostriker
[2004; see their eqs.~(1) and (3)] and we \emph{add} to the HMF the
subhalo mass function they have derived; we have checked that this
procedure does not alter substantially the overall mass density in
the galactic range. The subhalo MF by van den Bosch et al. (2005) is
extremely close to the Vale \& Ostriker one over the mass range of
interest here, and gives essentially indistinguishable results. To
account for effect (ii) we \emph{subtract} from the HMF the halo
mass function of galaxy groups and clusters. Estimates of the latter
obtained by different groups (Girardi \& Giuricin 2000; Martinez et
al. 2002; Hein\"am\"aki et al. 2003; Pisani et al. 2003; Eke et al.
2006) are in reasonable agreement for $M_h\ga 5\times 10^{12}\,
M_{\odot}$. At lower masses the galaxy group mass function is very
uncertain, and the recent study by Eke et al. (2006) finds a larger
abundance of low luminosity groups than previously reported from
smaller samples. On the other hand, from Fig.~8 (plus Fig.~15) of
Eke et al. (2006), it looks plausible that galaxies dominate the MF
for $M_h\la 5\times 10^{12}\, M_{\odot}$. We stress, however, that,
as discussed in the following, in the mass range considered here our
results are only weakly sensitive to whether or not the galaxy group
mass function is subtracted from the HMF.

\begin{figure}[t]
\plotone{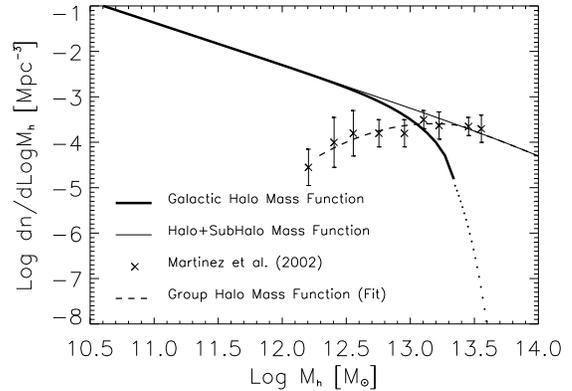} \caption{Galactic halo mass function. The
\textit{heavy solid} line shows the numerical results, obtained as
described in Sect.~\protect\ref{sect:ghmf}, which are very
accurately reproduced by the analytical approximation
[eq.~(\protect\ref{eq|GHMF})]. We have plotted the results by
Martinez et al. (2002) only for the halo mass range of interest
here. Beyond ${M_h}\approx 2\times 10^{13}\, M_{\odot}$, the halo
mass function is fully accounted for by groups and clusters of
galaxies; as a consequence the GHMF goes to zero (shown as a
\textit{dotted} line beyond this limit). }\label{fig|GHMF}
\end{figure}

The GHMF obtained subtracting from the HMF the group and cluster
mass function estimated by Martinez et al. (2002) is shown in
Fig.~\ref{fig|GHMF} and is well fit, in the range $11<\log
M_h/M_{\odot}<13.2$, by a Schechter function
\begin{equation}\label{eq|GHMF}
\mathrm{GHMF}(M_h)~\mathrm{d}M_h={\theta\over \tilde{M}}\,
\left({M_h\over \tilde{M}}\right)^{\alpha}\, e^{-(M_h/\tilde{M})}\,
\mathrm{d}M_h~,
\end{equation}
with $\alpha=-1.84$, $\tilde{M}=1.12\times 10^{13}\, M_{\odot}$
and $\theta=3.1\times 10^{-4}\,\hbox{Mpc}^{-3}$. The fall off at
high masses (where early-type galaxies dominate) mirrors the
increasing probability of multiple occupation of mass halos found
by Magliocchetti \& Porciani (2003) for $M\ga 3\times 10^{13}\,
M_{\odot}$ (see also Zehavi et al. 2005). Weak lensing
measurements also suggest an upper galaxy mass limit
$M_{\mathrm{max}} \la 3\times 10^{13}\, M_{\odot}$ (Kochanek \&
White 2001).

If two galaxy properties, $q$ and $p$, obey a monotonic
relationship we can write
\begin{equation}\label{eq|Jacobian}
\mathrm{\Phi}(p) {\mathrm{d}p\over\mathrm{d}q}\, \mathrm{d}q =
\mathrm{\Psi} (q)\, \mathrm{d}q~,
\end{equation}
where $\Psi(q)$ is the number density of galaxies with measured
property between $q$ and $q+\mathrm{d}q$ and $\Phi(p)$ is the
corresponding number density for the variable $p$. The solution is
based on a numerical scheme that imposes that the number of
galaxies with $q$ above a certain value $\bar{q}$ must be equal to
the number of galaxy halos with $p$ above $\bar{p}$ (see, e.g.,
Marinoni \& Hudson 2002; Vale \& Ostriker 2004), i.e.,
\begin{equation}\label{eq|criterion}
\int_{\bar{p}}^{\infty}\,\mathrm{\Phi}(p)\mathrm{d}p=\int_{\bar{q}}^{\infty}\,
\mathrm{\Psi}(q)\, \mathrm{d}q~.
\end{equation}
In the following $p\equiv M_h$ and $\Phi(p)\equiv
\mathrm{GHMF}(M_h)$, while the variable $q$ will be, in turn, the
luminosity, the stellar mass, the velocity dispersion, and the
central BH mass. It is worth noticing that in this way we establish
a direct link between the specific galaxy property and the halo mass
without any assumption or extrapolation of the DM density profile.

The monotonicity assumption is obviously critical for the
applicability of the present approach. However direct evidence of
monotonic relationships between several pairs of these quantities
has been reported (H\"aring \& Rix 2004; Ferrarese 2002; Baes et
al. 2003; Tremaine et al. 2002), and additional data supporting
the relationships derived here are discussed in the following.

We also implicitly assume that all galactic-size halos contain a
visible galaxy. This assumption underlies all major semi-analytic
models for galaxy formation, including the one by Granato et al.
(2004), that we adopt as our reference model. The successes of this
model in reproducing the redshift-dependent galaxy luminosity
function in different wavebands provide strong support to this view.

\begin{figure}[t]
\plotone{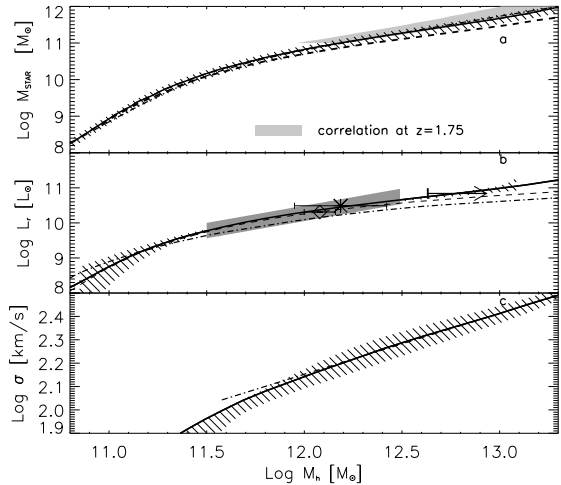} \caption{($a$) Mass in stars versus halo mass. The
thick \textit{solid} line shows the numerical results while the
\textit{dot-dashed} line (difficult to distinguish from the solid
line) represents the analytic fitting formula
[eq.~(\protect\ref{eq|MsMh})]. The \textit{dashed} line has been
obtained using the GSMF by Bell et al. (2003); the \textit{barred}
area represents the uncertainty associated to the mass to light
ratio; the \textit{shaded} area illustrates the result at $z=1.75$,
based on data by Fontana et al. (2004). ($b$) $r^*$-band luminosity
as a function of halo mass. Again, the thick \textit{solid} line
shows the numerical results and the analytic fitting formula
[eq.~(\protect\ref{eq|LrMh})] is indistinguishable from it. The
\textit{barred} area represents the uncertainty associated to the LF
(dominant at the low-mass end) and to the GHMF (dominant at the
high-mass end); the thin \textit{dashed} line is the numerical
result obtained without removing the galaxy groups and clusters from
the HMF; the \textit{dot-dashed} line is the result by Vale \&
Ostriker (2004); the \textit{shaded} region shows the result by
Kleinheinrich et al. (2004); the \textit{arrow} is from O'Sullivan
\& Ponman (2004); the \textit{diamond} is from Hoekstra et al.
(2004); the \textit{star} is from Guzik \& Seljak (2002). ($c$)
$\sigma$--$M_h$ relation. As before, the thick \textit{solid} line
shows the numerical results while the \textit{dot-dashed} line shows
the analytic formula [eq.~(\protect\ref{eq|sigmaMh})] holding in the
large $M_h$ limit. The \textit{barred} area represents the
uncertainty. }\label{fig|MsLrMh}
\end{figure}

The result for the stellar mass, obtained setting
$\mathrm{\Psi}(q)=\mathrm{GSMF}(M_{\mathrm{star}})$
[eq.~(\ref{eq|GSMF})] is plotted in Fig.~\ref{fig|MsLrMh}$a$. We
find that its relationship with the host halo mass is well
approximated by
\begin{equation}\label{eq|MsMh}
M_{\mathrm{star}}\approx 2.3\times 10^{10}\, M_{\odot}\, \frac {
(M_h/3\times 10^{11}\, M_{\odot} )^{3.1} }{1+ (M_h/3\times
10^{11}\, M_{\odot})^{2.2}} ~.
\end{equation}
The calculations  for the baryonic mass is strictly analogous.
Setting $\mathrm{\Psi}(q)= \phi(L_{r})$ [eq.~(\ref{eq|LF})] we
also derived the approximated behavior of the luminosity as a
function of the halo mass
\begin{equation}\label{eq|LrMh}
L_{r}\approx 1.2\times 10^{10}\, L_{\odot}\, \frac {(M_h/3\times
10^{11}\, M_{\odot} )^{2.65} }{1+ (M_h/3\times 10^{11}\,
M_{\odot})^{2.00}}~,
\end{equation}
and of the halo mass as a function of luminosity
\begin{equation}\label{eq|MhLr}
{M_h\over 3\times 10^{11}\, M_{\odot}} = \left[\left(\frac
{L_{r}}{1.3\times 10^{10}\, L_{\odot}}\right)^{0.35}+\left(\frac
{L_{r}}{1.3\times 10^{10}\, L_{\odot}}\right)^{1.65}\right]~.
\end{equation}
Both stellar mass and luminosity exhibit a double power law
dependence on halo mass with a break around
$M_{h,{\mathrm{break}}}\sim 3\times 10^{11}\,M_{\odot}$,
corresponding to a luminosity $L_{r, {\mathrm{break}}}\sim 6\times
10^9\,L_{\odot}$.

The derivation of the $L_r$--$M_h$ relation is quite sensitive to
uncertainties in the LF and in the GHMF. In the range $10^9\,
L_{\odot}\la L_r\la 10^{11}\, L_{\odot}$ the LF is rather
precisely known (see the comparison of different LF estimates in
Fig.~1 of Shankar et al. 2004), while the effect of uncertainties
in the GHMF becomes significant at large masses. On the whole, the
$L_r$--$M_h$ relation is quite accurately determined in the range
$10^{11}\, M_{\odot}\la M_h\la 10^{13}\, M_{\odot}$
(Fig.~\ref{fig|MsLrMh}$b$). At low luminosities the errors on the
LF rapidly increase, becoming a factor of about $2$ for $L_r\la
3\times 10^8\, L_{\odot}$. In order to illustrate the consequences
on the $L_r$--$M_h$ relation, we have computed it using the
$1\,\sigma$ upper and lower boundaries of the LF by Nakamura et
al. (2003). In the former case, the low-luminosity portion of the
$L_r$--$M_h$ relation flattens from a slope of $\sim 2.6$ to $\sim
1.9$; in the latter case, it decreases almost exponentially.
Therefore, the extrapolation of the above relationships to $L_r\ll
10^9\, L_{\odot}$ and, correspondingly, to $M_h\ll 10^{11}\,
M_{\odot}$ must be taken cautiously. Of course, the same
conclusion holds for relationships involving other statistics of
galaxies and of their host halos. This emphasizes the need of a
precise determination of the low luminosity end of the LF.

In Fig.~\ref{fig|MsLrMh}$b$ our estimate of the $L_r$ vs. $M_h$
relation is compared with observational results for galaxies whose
$M_h$ could be derived based on two different methods: (i) an
X-ray-based mass model of the isolated elliptical galaxy NGC 4555
(O'Sullivan \& Ponman 2004), in which the gravitational potential is
known up to about $1/8$ of the virial radius (the mass within this
radius is shown as a lower limit in Fig.~\ref{fig|MsLrMh}$b$); (ii)
weak-lensing observations that provide the shear field around a
number of galaxies of average luminosity $L$, from which it is
possible to infer the projected mass density and eventually to
extrapolate the virial mass by assuming a DM profile (Guzik \&
Seljak 2002; Kleinheinrich et al. 2004; Hoekstra et al. 2004). In
Fig.~\ref{fig|MsLrMh}$b$ we also show for comparison the $L_r$ vs.
$M_h$ relation obtained from eq.~(\ref{eq|Jacobian}) setting
$\Phi(p)=HMF(M_h)$. It is worth noticing that if the group and
cluster MF is not subtracted from the HMF, our results vary only at
high masses, and the changes do not exceed 0.2 dex up to $M_h \simeq
2\times 10^{13}\, M_{\odot}$.

\begin{figure}[t]
\plotone{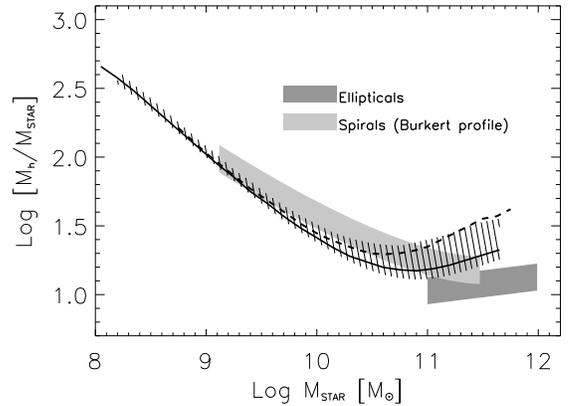} \caption{Halo to stellar mass ratio as a function
of the stellar mass.  The \textit{solid} line is the result of
numerical calculations using eq.~(\protect\ref{eq|criterion}) with
$\Phi(p)\equiv \mathrm{GHMF}(M_h)$ and $\mathrm{\Psi}(q)\equiv
\mathrm{GSMF}(M_{\mathrm{star}})$, as given by
[eq.~(\protect\ref{eq|GSMF})]; the \textit{barred} area represents
the uncertainty associated to the mass to light ratio. The
\textit{dashed} line has been obtained using the GSMF by Bell et al.
(2003). The Cole et al. (2001) GSMF with a Salpeter (1955) IMF
yields results very close to ours. The \textit{dark shaded} area
represents the data on giant elliptical galaxies by Gerhard et al.
(2001); the \textit{light shaded} area represents data on spiral
galaxies by Persic, Salucci \& Stel (1996) and Salucci \& Burkert
(2000).}\label{fig|Ratio}
\end{figure}

As a further check,  our estimate of the ratio
$M_{h}/M_{\mathrm{star}}$ as a function of $M_{\mathrm{star}}$, is
compared in Fig.~\ref{fig|Ratio} with estimates derived by
extending to the virial radius the inner mass models of a number
of giant ellipticals (Gerhard et al. 2001) and spirals (Persic,
Salucci \& Stel 1996; Salucci \& Burkert 2000). We stress that
these results require extrapolations to the virial radius of the
density profile, assumed to have the Navarro, Frenk \& White
(1996) shape, while our estimate does not need any assumption on
DM density profile. It is apparent that these independent results
are in nice agreement with our findings.

The dependence of the luminosity on the halo mass has been
investigated also by Vale and Ostriker (2004); their result is
also shown in Fig.~\ref{fig|MsLrMh}$b$. They exploited the 2dF
galaxy luminosity function in the $b_J$-band (Norberg et al.
2002), extrapolating it beyond the range of magnitudes where it
was defined. The difference in the $L_r$-$M_h$ relation between
our estimate and theirs is due to the steeper slope of the LF
adopted by them. There is also a small difference in the
normalization of the LF, but this is of minor importance.

At high masses the direct comparison of the galaxy LF  to the halo
plus subhalo number density [cf. their eq.~(9)] results in a
slight flattening of the relation (dot-dashed curve for Vale \&
Ostriker, dashed curve for our calculations). As Vale and Ostriker
(2004) pointed out, in this way the mass term refers to the entire
halo hosting the group or the cluster and not to just the galaxy
halo.

Guzik \& Seljak (2002) modelled the galaxy-galaxy lensing trying
to separate the central galactic contributions from contributions
of the surrounding groups and clusters. Their model applied to the
SDSS data on galaxy lensing yields $M_h/L_r\approx 50\,
M_{\odot}/L_{\odot}$ at the characteristic luminosity $3\times
10^{10}\, L_{\odot}$ for early-type galaxies, in keeping with  our
results. They also found a luminosity dependence $M_h\propto
L^{1.4\pm 0.2}$, compatible with the high-luminosity slope of
eq.~(\ref{eq|MhLr}). However, we find that, at low luminosities,
the slope significantly flattens toward a dependence $M_h\propto
L^{0.35}$, while Guzik \& Seljak (2002) assume a single power law
relation.

Van den Bosch et al. (2003, 2005) computed the conditional
luminosity function of early- and late-type  galaxies, a
statistics linking the distribution of galaxies to that of the DM.
They concluded that the MLR has a minimum $M_h/L_{r}\sim 45-70\,
M_{\odot}/L_{\odot}$ at $M_h\sim 2-4 \times 10^{11}\, M_{\odot}$,
consistent with eq.~(\ref{eq|MhLr}).

Marinoni \& Hudson (2002) investigated the MLRs of the virialized
systems, which include galaxies, groups and clusters. By comparing
their luminosity function to the $\Lambda$CDM halo mass function,
they concluded that the MLR has a broad minimum around $L_B\approx 3
\times 10^{10}\, L_{\odot}$. The slopes at low and high luminosity
are $-0.5$ and $+0.5$, respectively. Our slope is similar at low
masses, where practically all virialized systems are galaxies and
thus the comparison is meaningful. The studies by Eke et al. (2004,
2006) of the variation of the MLR with size of galaxy groups is
fully consistent with a minimum at approximately the same
luminosity.

By comparing the HMF and the LF, as we have done for local
galaxies, it is possible to infer the $M_{\mathrm{star}}$--$M_{h}$
relation even at substantial redshifts. For the GSMF we use a
simple linear fit [$\log(\phi(M_{\rm star})/\hbox{Mpc}^3) = -1.7
\log(M_{\rm star}/M_\odot) +16.1$, holding for $11\le \log(M_{\rm
star}/M_\odot)\le 12$] to the data by Fontana et al. (2004) at
$\bar{z}=1.75$, and we approximate the GHMF with the HMF computed
at the same redshift. The result, shown by the shaded area in
Fig.~\ref{fig|MsLrMh}$a$, has to be taken as an upper limit since
we have neglected the contribution of galaxy groups to the HMF.
Clearly, our estimate becomes increasingly uncertain as we
approach the upper limit of the interval where GSMF is
observationally estimated; this is reflected in the increased
width of the shaded area. Nevertheless, the
$M_{\mathrm{star}}$--$M_h$ relation at $\bar{z}=1.75$ turns out to
be quite close to the local one, indicating that for large
galaxies the $M_{\mathrm{star}}$--$M_{h}$ relation was already in
place at redshift $z\ga 1$ in line with the anti-hierarchical
baryon collapse scenario developed by Granato et al. (2001; 2004).

Sheth et al. (2003) estimated the Velocity Dispersion Function
(VDF) of spheroidal galaxies using a sample drawn from the SDSS
survey and have built a simple model for the contribution to the
VDF of the bulges of spirals, which dominate at low velocity
dispersions. Their estimate covers the range $80$ km s$^{-1}\leq
\sigma\leq 400$ km s$^{-1}$. Comparing the global VDF (including
both early and late type galaxies, as shown in Fig.~6 of Sheth et
al.), and the GHMF with the same technique presented above [cf.
eqs.~(\ref{eq|Jacobian}) and (\ref{eq|criterion})], we can derive
the $\sigma$--$M_h$ relationship shown in panel (c) of
Fig.~\ref{fig|MsLrMh}. For $M_h\geq 6.3\times 10^{11}$, the
relationship is accurately represented by a simple power-law:
\begin{equation}\label{eq|sigmaMh}
\sigma \approx 110 \, \mathrm{km\, s}^{-1}\, \left({M_h\over
6.3\times 10^{11}\, M_{\odot}}\right)^{1/3}~,
\end{equation}
while it steepens for lower halo masses.  The uncertainties
strongly increase for $\sigma\la 80$ km s$^{-1}$, corresponding to
$M_h\la 10^{11}\, M_{\odot}$. Note that these relationships must
break down in the low-$\sigma$ (and low-$M_h$) regime. In fact,
the close match found by Cirasuolo et al. (2005) between the VDF
and the virial velocity function (derived from the halo mass
distribution function integrated over redshift) indicates that
halos more massive than $M_h\sim 10^{11}\, M_{\odot}$ are
generally associated to spheroidal galaxies or to later type
galaxies with bulge velocity dispersions $\sigma\ga 80$ km
s$^{-1}$. On the other hand, the fraction of galactic halos
associated to essentially bulgeless late-type galaxies must
increase with decreasing $M_h$, so that the integral of the VDF
deviates from that of the GHMF and eq.~(\ref{eq|criterion}) no
longer applies.

\begin{figure}[t]
\plotone{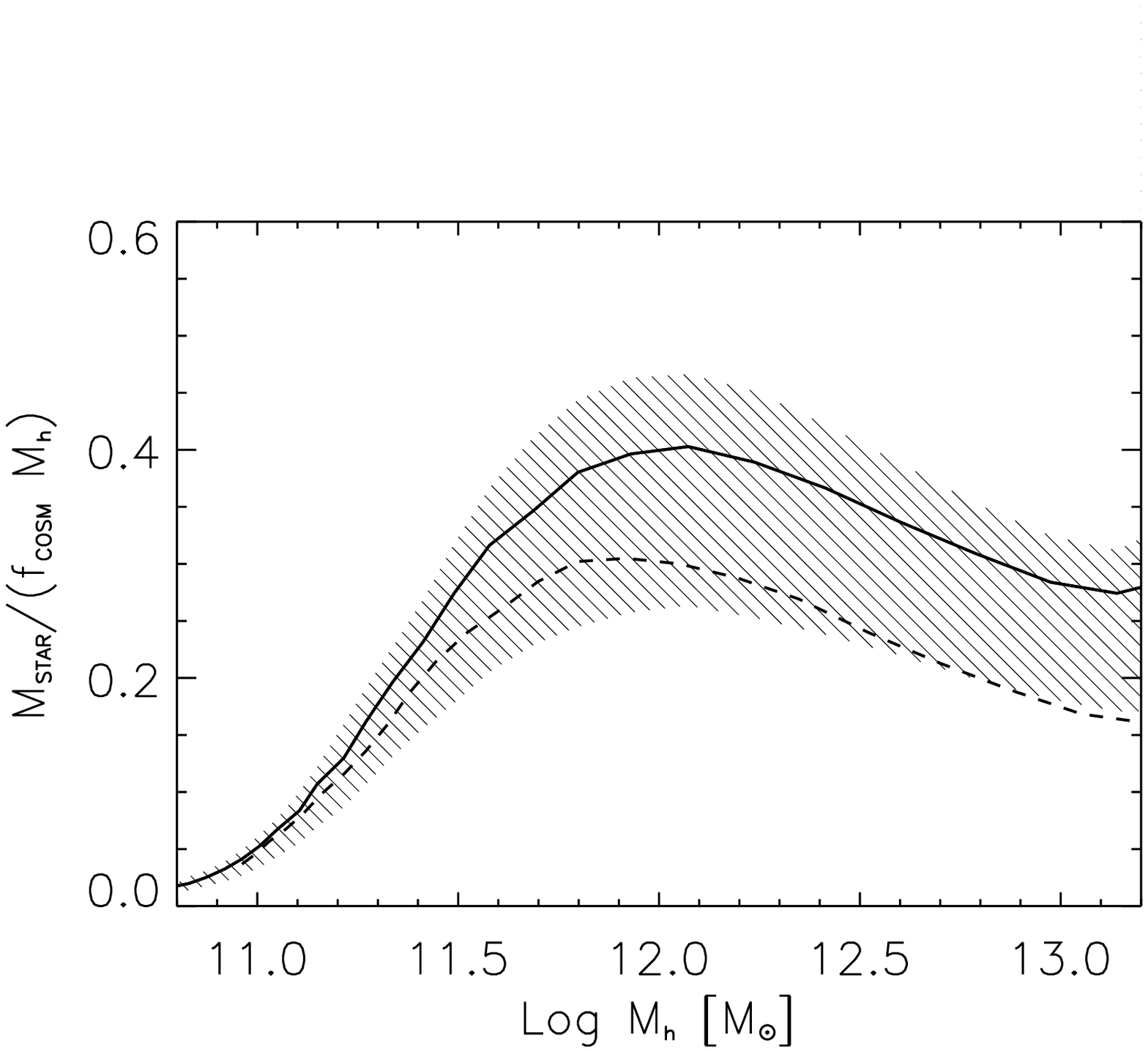} \caption{Fraction of primordial gas turned into
stars as a function of halo mass. The \textit{solid} line has been
obtained numerically from eq.~(\protect\ref{eq|criterion}) with
$\Phi(p)\equiv \mathrm{GHMF}(M_h)$ and $\mathrm{\Psi}(q)\equiv
\mathrm{GSMF}(M_{\mathrm{star}})$, as given by
[eq.~(\protect\ref{eq|GSMF})]; the \textit{barred} area represents
the uncertainty associated to the mass to light ratio. The
\textit{dashed} line has been obtained using the GSMF by Bell et al.
(2003). We have set $f_{\mathrm{cosm}}= 1/6$.
}\label{fig|Efficiency}
\end{figure}

Figure~\ref{fig|Efficiency} shows the ratio of the mass in stars to
the initial baryon mass associated with each halo, assumed to be
$M_{b,i}=f_{\mathrm{cosm}} M_h$, as a function of $M_h$. It
illustrates the ``inefficiency'' of galaxies, especially of those of
low halo mass, in retaining baryons. As discussed in
\S\,\ref{sect:feed}, the shape of the $M_{\rm star}/f_{\rm cosm}
M_h$ can be understood in terms of feedbacks: at low masses the SN
feedback is very efficient in removing the gas, thus quenching the
star formation; moving toward higher masses (for $\log(M_{\rm BH})
\ga 7.5$, corresponding to $\log(M_h \ga 12)$) the AGN feedback
becomes more and more powerful, to the point of sweeping out most of
the initial baryons.

Our result is at odds with the claim by Guzik \& Seljak (2002) of
a high efficiency, up to $75\%$, in turning primordial gas into
stars. However, the claim is based on a MLR $M_h/L_i \approx 17
h\, M_{\odot}/L_{\odot}$ in the $i'$ band for late-type galaxies,
a factor of 3 lower than the value found for the early-type ones.
As the authors themselves point out, the statistical significance
of this result is marginal, due to the weak lensing signal for the
fainter late-type galaxy sample. {The GSMFs by Cole et al. (2001)
and Bell et al. (2003) yield similar efficiencies, which are very
close to our estimates for relatively low halo masses, but lower
for large masses, yet within the estimated uncertainties.}

\section{Black hole vs. halo mass}

\begin{figure}[t]
\plotone{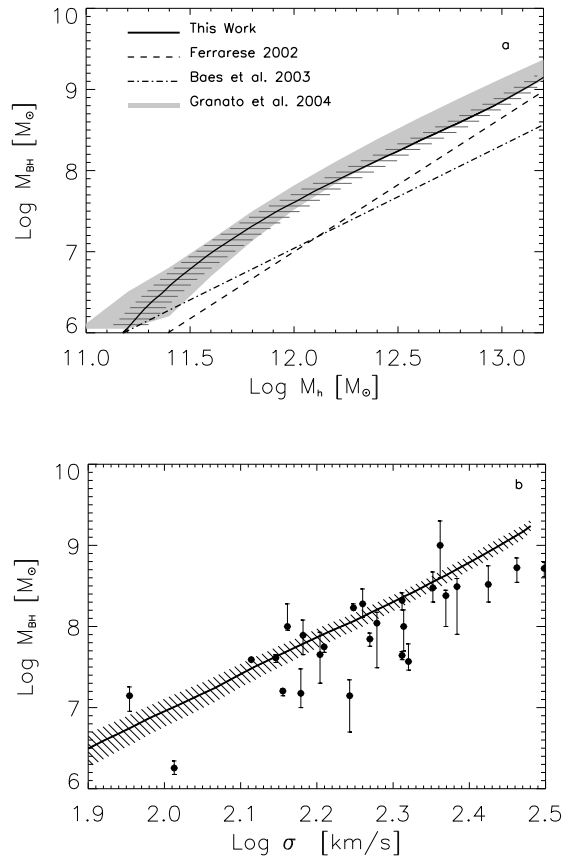} \caption{Upper panel: super-massive BH mass versus
halo mass. The \textit{solid} line has been obtained numerically
from eq.~(\protect\ref{eq|criterion}) with $\Phi(p)\equiv
\mathrm{GHMF}(M_h)$ and $\mathrm{\Psi}(q)$ is the local
super-massive BH mass function estimated by Shankar et al. (2004).
The \textit{dashed} line is the relation by Ferrarese et al. [2002;
her eq.~(6)]; the \textit{dot-dashed} line is the relation by Baes
et al. (2003). The \textit{shaded} area represents the prediction of
the Granato et al. (2004) model. Lower panel: $M_{\rm BH}$--$\sigma$
relation obtained combining the $M_{\rm BH}$--$M_h$ relation (upper
panel) with the $\sigma$--$M_h$ relation (panel (c) of
Fig.~\protect\ref{fig|MsLrMh}). The data are from Ferrarese \& Ford
(2005). In both panels the \textit{barred} area reflects the
uncertainty associated to the BH Mass Function. }\label{fig|MbhMh}
\end{figure}

The relation between the central supermassive black hole and the
halo mass $M_{h}$ is relevant in the framework of theories for the
origin and evolution of both galaxies and AGNs (Silk \& Rees 1998;
Monaco, Salucci \& Danese 2000; Granato et al. 2001; Ferrarese
2002; Granato et al. 2004). To constrain such relation we adopted
the procedure presented in the previous Section
[eqs.~(\ref{eq|Jacobian}) and (\ref{eq|criterion})], replacing the
function $\Psi(q)$ with the local BH mass function (Shankar et al.
2004). We assume that each galactic halo hosts just one
supermassive BH. Our result is shown in Fig.~\ref{fig|MbhMh},
where the barred area illustrates the errors due to the
observational uncertainties on the BH mass function, as estimated
by Shankar et al. (2004). We find good agreement, within the
estimated uncertainties, with the predictions of the Granato et
al. (2004) model.

The relationship can be approximated by
\begin{equation}\label{eq|MbhMh}
M_{\rm BH}\approx 6\times 10^{6} \, M_{\odot}\, \frac
{(M_h/2.2\times 10^{11}\, M_{\odot} )^{3.95} }{1+ (M_h/2.2\times
10^{11}\, M_{\odot})^{2.7}}~.
\end{equation}
Again a double power law with a break at $M_h\sim 3\times 10^{11}\,
M_{\odot}$ is a very good representation of our results. At the high
mass end, the BH mass is nearly proportional to the halo mass
($M_{\rm BH}\propto M_h^{1.25}$), while at low masses the relation
steepens substantially ($M_{\rm BH}\propto M_{h}^{3.95}$).

In Fig.~\ref{fig|MbhMh} we also compare our estimate of the
$M_{\rm BH}$--$M_h$ relation with that of Ferrarese (2002), who
first investigated this issue from an observational point of view.
She derived a power-law relationship between the bulge velocity
dispersion and the maximum circular velocity, $v_c$, for a sample
of spiral and elliptical galaxies spanning the range $100\la
v_c\la 300$ km s$^{-1}$, and combined it with the relationship
between $v_c$ and the virial velocity, $v_{\rm vir}$ based on the
numerical simulations by Bullock et al. (2001) and with the
$M_h$--$v_{\rm vir}$ relationship given by the $\Lambda$CDM model
of the latter authors for a virialization redshift $z_{\rm
vir}\sim 0$, to obtain a $M_h$--$\sigma$ relation. Coupling it
with one version of the observed BH mass vs. stellar velocity
dispersion relationship ($M_{\rm BH}\propto \sigma^{4.58}$) she
obtained $M_{\rm BH} \propto M_{h}^{\alpha}$, with $\alpha=
1.65$--1.82. Baes et al. (2003) with the same method, but assuming
$M_{\rm BH}\propto \sigma^{4.02}$ and with new velocity dispersion
measurements of spiral galaxies with extended rotation curves,
yielding a slightly different $v_c$-$\sigma$ relation, found
$M_{\rm BH} \propto M_{h}^{1.27}$. It is apparent from
Fig.~\ref{fig|MbhMh} that our result differs substantially in
normalization, while the high-mass slope is remarkably close to
that obtained by Baes et al. (2003). It should be noted that the
$M_h$--$V_{\rm vir}$ relation depends on the virialization
redshift. For $z_{\rm vir} \simeq 3$ [the median virialization
redshift for galaxies with velocity dispersions in the range
probed by Ferrarese and Baes et al., according to the analysis by
Cirasuolo et al. (2005; see their Fig.~1)], its coefficient would
be a factor of $\simeq 4.25$ lower than that used by Ferrarese
(2002) and Baes et al. (2003) and the coefficients of the $M_{\rm
BH} \propto M_{h}$ relations would be larger by a factor of
$\simeq 5.6$ in the case of eq.~(6) of Ferrarese(2002) or of
$\simeq 4$ in the case of Baes et al. (2003),  bringing them much
closer to ours. In fact, as suggested by Loeb \& Peebles (2003)
and shown in details by Cirasuolo et al. (2005), the velocity
dispersion of the old stellar population (whose mass is related to
the central BH mass) is closely linked to halo mass and
characteristic velocity at the virialization redshift.

As a consistency test, we have combined the $M_{\rm BH}$--$M_h$
relation, shown in the upper panel of Fig.~\ref{fig|MbhMh}, with
the $\sigma$--$M_h$ relation, shown in panel (c) of
Fig.~\ref{fig|MsLrMh}, to obtain the $M_{\rm BH}$--$\sigma$
relation (see Ferrarese \& Ford 2005 for a review), which turns
out to be consistent with the data, as shown by the lower panel in
Fig.~\ref{fig|MbhMh}.

\section{Feedback from stars and AGNs}\label{sect:feed}

The dependence of the star and BH masses on the halo mass found in
the previous Section, suggests that different physical mechanisms
are controlling the star formation  and the BH growth above and
below $M_{h,{\mathrm{break}}} \sim 3 \times 10^{11}\, M_{\odot}$,
corresponding to $M_{\mathrm{star}}\sim 1.2\times 10^{10}\,
M_{\odot}$ after eq.~(\ref{eq|MsMh}), and to $L_r\sim 5 \times
10^9\, L_{\odot}$ (or $M_r\sim -19.6$) after eq.~(\ref{eq|LrMh}).
It is worth noticing that the analysis of a huge sample of
galaxies drawn from the SDSS shows that around
$M_{\mathrm{star}}\approx 2$--$3\times 10^{10}\, M_{\odot}$ and
$M_r\sim -19.8$ there is a sort of transition in the structure and
stellar ages of galaxies (Kauffmann et al. 2003; Baldry et al.
2004).

The efficiency of star formation within galactic halos of different
masses is the result of several processes. The most important are:
(i) the cooling of the primordial gas within the virialized halos
(White \& Rees 1978); (ii) the injection of large amounts of energy
into the ISM by supernova explosions (Dekel \& Silk 1986; White \&
Frenk 1991; Granato et al. 2001; Romano et al. 2002)  and by the
central quasar (Silk \& Rees 1998; Granato et al. 2001, 2004; Lapi
et al. 2005). All these processes have been implemented in the model
of Granato et al. (2004). A simplified, analytical rendition of this
model is presented in the Appendix A.

\begin{figure}[t]
\plotone{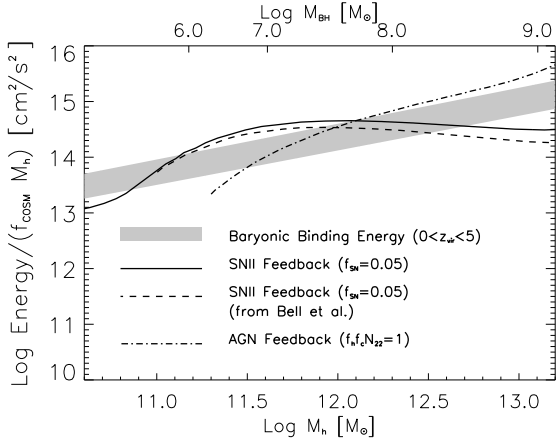} \caption{Specific energy feedback from stars and
AGNs compared to the baryon specific binding energy within the host
halo, as a function of the halo mass. The \textit{solid} line refers
to our estimate of the SN feedback while the \textit{dashed} line is
the SN feedback obtained using the GSMF of Bell et al. (2003). The
\textit{dot-dashed} line is our estimate of the AGN feedback. The
\textit{shaded} area shows the specific binding energy of the gas in
the DM potential well, for virialization redshifts $0\leq
z_{\mathrm{vir}}\leq 5$.}\label{fig|feedback}
\end{figure}

The impact of stellar and AGN feedback is illustrated by
Fig.~\ref{fig|feedback}. The binding energy of baryons in the DM
potential well per unit baryonic mass as a function of the halo
mass [cf. eq.~(\ref{eq|binding})] for $0\leq z_{\mathrm{vir}}\leq
5$ is shown by the shaded area.  To compute the overall energy per
unit baryonic mass injected in the gas by supernovae
($E_{\mathrm{star}}$) and by AGNs ($E_{\rm AGN}$), we have
exploited eqs.~(\ref{eq|stellarfeed}) and (\ref{eq|AGNfeed}),
respectively, where $M_{\mathrm{star}}$ and $M_{\rm BH}$ as
functions of the halo mass are given by eqs.~(\ref{eq|MsMh}) and
(\ref{eq|MbhMh}), respectively. Then we divided the overall
energies by the initial baryon mass $M_{b,i}=f_{\mathrm{cosm}}
M_h$. Figure~\ref{fig|feedback} shows that for large masses the
gas can be efficiently removed by the AGN feedback which
overwhelms the binding energy. For small masses the supernova
feedback dominates but appears to be insufficient to remove the
gas associated to the host halo, due to the above mentioned
problem that the observed $M_{\mathrm{star}}$--$M_h$ relation
inferred from the data exhibits a too steep low-mass slope. The
flattest slope allowed by the data, discussed in
\S$\,$\ref{sect:ghmf}, would largely alleviate, but not completely
overcome, this problem.

The relative importance of the two feedbacks depends on their
efficiency in transferring the available energy to the gas. It is
interesting that with the efficiencies used in
Fig.~\ref{fig|feedback} the crossing point is quite close to
$M_{h,{\mathrm{break}}}\approx 3\times 10^{11}\,M_{\odot}$. As
discussed by Granato et al. (2004) and Cirasuolo et al. (2005), a
more accurate evaluation of the efficiencies can be obtained by
fitting statistics of galaxies and of AGNs, such as LFs at low and
high redshift, the Faber and Jackson relation, and the local BH
mass function.

A more quantitative insight into the role of the key ingredients of
the model is provided by the analytic calculations presented in the
Appendix A. So long as the star formation rate obeys
eq.~(\ref{eq|dotMc}), the mass in stars at a the present time $t$,
assumed to be $>>t_c$, is given, after eq.~(\ref{eq|Ms}), by:
\begin{equation}\label{eq|MsMhfeed}
M_{\mathrm{star}} \propto f_{\mathrm{surv}}\,
{f_{\mathrm{cosm}}M_{h}\over 1-R+\alpha}~,
\end{equation}
where $f_{\mathrm{surv}}$ is the fraction of stars that survive up
to now.

For large halo masses, where the stellar feedback is less efficient
($\alpha \la 1$), the quantity $1-R+\alpha$ is a slowly decreasing
function of the halo mass, so that $M_{\mathrm{star}}$ is
approximately proportional to $M_h$. However, in this case, the
fraction of gas turned into stars is controlled by the AGN feedback,
which, as shown by the full treatment by Granato et al. (2004), for
$M_h \ga 3 \times 10^{11}\, M_\odot$ expels an approximately
constant fraction of the initial gas, thus preserving the
approximate proportionality between $M_{\mathrm{star}}$ and $M_h$,
in agreement with eq.~(\ref{eq|MsMh}).

The effective optical depth, which rules the flow of the cold gas
into the reservoir around the BH [cf. eq.~(\ref{eq|dotMres})], is
large ($\tau \ga 1$) for large galaxies, implying, after
eq.~(\ref{eq|Mbh}), $M_{\rm BH}\approx 1.2\times 10^{-3}\,
M_{\mathrm{star}}$. As a consequence, in the high mass limit, the BH
mass must have a dependence on the halo mass very similar to that of
the stellar mass, in agreement with eq.~(\ref{eq|MbhMh}).

For $M_h \ll 10^{12}\, M_{\odot}$ the dominant term in the
denominator in the right-hand side of eq.~(\ref{eq|MsMhfeed}) is
the effective efficiency of the SN feedback, $\alpha \propto
M_{h}^{-2/3}$ [cf. eq.~(\ref{eq|SNeff})]. Therefore we get
\begin{equation}\label{eq|slopeMsMh}
M_{\mathrm{star}}\propto f_{\mathrm{surv}}\, M_h^{5/3}~.
\end{equation}
This limiting dependence has been derived theoretically also by
Dekel \& Woo (2003) with similar assumptions. On the other hand,
such a relation is significantly flatter than that inferred from the
data [cf. eq.~(\ref{eq|MsMh}) and Fig.~\ref{fig|feedback}]. Possible
explanations may be that in less massive halos the initial baryon
fraction is lower by effect of reheating of the intergalactic
medium, hindering the infall of baryons into the shallower potential
wells, or that the SN efficiency in removing the gas is higher.
However, the difference must not be overemphasized, in view of the
large uncertainties on the shape of the $M_{\mathrm{star}}$--$M_h$
and $M_h$-$L_r$ relations at low masses/luminosities, induced by our
poor knowledge of the low-luminosity portion of the LF. As discussed
in \S~\ref{sect:ghmf}, the data are consistent also with a flatter
relation ($M_{\mathrm{star}}\propto M_h^{1.9}$).

Since in the mass range $M_h\la  10^{11}\, M_{\odot}$ the optical
depth is small ($\tau \ll 1$), from eqs.~(\ref{eq|Mbh}) and
(\ref{eq|optdepth}) we obtain:
\begin{equation}\label{eq|slopeMbhMh}
M_{\rm BH}\propto M_{\mathrm{star}}\, \tau \propto M_h^{7/3}~.
\end{equation}
Thus this simple model predicts that the low mass slope of the
$M_{\rm BH}$--$M_h$ relation is steeper than that of the
$M_{\mathrm{star}}-M_h$ relation, because of the decrease of the
optical depth with mass $\tau\propto M_h^{2/3}$, entailing a lower
capability of feeding the reservoir around the BH. Interestingly,
a steepening by approximately this amount is also found from our
analysis of observational data [cf. eqs.~(\ref{eq|MsMh}) and
(\ref{eq|MbhMh})].

\section{Conclusions}

We computed the stellar and baryon mass function in galaxies
exploiting $M/L$ ratios for stars and gas derived from galaxy
kinematics. The results turn out to be in agreement with previous
analyses based on stellar population models. The total baryonic mass
density in galactic structures amounts to $\Omega_b^G\approx (3.6\pm
0.7)\times 10^{-3}$, of which $\sim 40\%$ resides in late-type
galaxies. This result confirms the well-known conclusion that only a
fraction $\la 10\%$ of the cosmic baryons are presently in stars and
cold gas within galaxies.

The present-day galaxy halo mass function, i.e., the number density
of halos of mass $M_h$ containing a single baryonic core, has been
estimated by adding the subhalos to the halo mass function and by
subtracting the contributions from galaxy groups and clusters. Such
subtraction, which  is required to single out galactic halos, has
however a minor effect over the mass range of interest here.

Approximated analytic relationships between the halo mass, $M_h$,
the mass in stars, $M_{\mathrm{star}}$,  and the $r^*$-band
luminosity, $L_r$, have been obtained from the functional
equations $\mathrm{\Phi}[>M_h(q)]= \mathrm{\Psi}(>q)$, where
$\mathrm{\Phi}(>M_h)$ is the number density of galactic halos
larger that $M_h$ and $\mathrm{\Psi}(>q)$ is the number density of
galaxies with either stellar mass greater than $M_{\mathrm{star}}$
or luminosity greater than $L_r$. The results are in good
agreement with $M_h/L_r$ ratios inferred through X-ray mapping of
the gravitational potential and through gravitational lensing.
Both relations exhibit a double-power law shape with a break
around $M_{h,{\mathrm{break}}}\approx 3\times 10^{11}\,
M_{\odot}$, corresponding to $M_{\mathrm{star},\mathrm{break}}
\approx 1.2\times 10^{10}\, M_{\odot}$ and to an absolute
magnitude $M_{r,{\mathrm{break}}}\approx -19.6$. A transition at
about the same magnitude in the galaxy properties has been
evidenced by Kauffmann et al. (2003) and Baldry et al. (2004).

An additional interesting outcome of our analysis is that the
$M_{\mathrm{star}}$--$M_h$ relation is already established at
redshift $z\approx 1.7$, in line with the theoretical expectation
of the anti-hierarchical baryon collapse scenario (Granato et al.
2004).

Applying the same technique to the local velocity dispersion
function of galaxies and to the black hole mass function, we have
also computed the $\sigma$--$M_h$ and $M_{\rm BH}$--$M_h$
relationships. The former is quite close to a single power law
$\sigma \propto M_h^{1/3}$. The latter is again a double power law
breaking approximately at $M_{h,{\mathrm{break}}}$, corresponding
to $M_{\rm BH}\sim 9\times 10^{6}\, M_{\odot}$. The associated
velocity dispersion, $\sigma \simeq 88$ km s$^{-1}$, is very close
to the first estimate of the critical velocity dispersion for the
gas removal by SN explosions given by Dekel \& Silk (1986), who
found a critical halo velocity $V_{\mathrm{crit}}\sim 120$ km
s$^{-1}$, corresponding to a critical velocity dispersion
$\sigma_{\mathrm{crit}}\sim 80\,\hbox{km}\,\hbox{s}^{-1}$.

As a test of our results, we combined the $M_{\rm BH}$--$M_h$
relation [eq.~(\ref{eq|MbhMh})] with the $\sigma$--$M_h$ relation
[eq.~(\ref{eq|sigmaMh})]; as shown by the lower panel of
Fig.~\ref{fig|MbhMh}, the resulting $M_{\rm BH}$--$\sigma$
relation is consistent with the observational data.

The relationships we have obtained are model-independent and can
be interpreted in terms of feedback effects by supernovae and AGNs
in galactic structures. We also presented a simple feedback model,
which nicely reproduces the approximate proportionalities $M_{\rm
BH}\propto M_{\mathrm{star}}\propto M_h$ observed in the high mass
range, and the break of these relationships at
$M_{h,{\mathrm{break}}}\approx 3\times 10^{11}\, M_{\odot}$.

At low masses, the $M_{\mathrm{star}}$--$M_h$ relation derived
here ($M_{\mathrm{star}}\propto M_h^{3.1}$) is steeper than that
yielded by the model ($M_{\mathrm{star}}\propto M_h^{5/3}$), and
would imply that only a tiny fraction of the baryons initially
associated with the halo remains within it in the form of stars
(and we know that the gas does not add much to the baryon content
of galaxies). On the other hand, if the amount of stars formed is
so low, for a standard Salpeter IMF the energy injected by SN
explosions is insufficient to expel the residual gas if the baryon
fraction is close to the cosmic value. Thus, if the slope of the
$M_{\mathrm{star}}$--$M_h$ relation really is as steep as its face
value suggests, we must conclude that either the initial baryon
fraction in low-mass galaxies was substantially lower than the
cosmic value (due, e.g., to a pre-heating of the intergalactic
medium hampering the infall of baryons into shallow potential
wells) or that the SN feedback in these objects was substantially
stronger than in more massive galaxies. As discussed in
\S$\,$\ref{sect:ghmf}, however, the uncertainties on the low
luminosity portion of the LF are large enough to allow for a
flatter slope, closer to the model prediction and almost
sufficient to grant the gas removal by SN feedback.

%
The errors shown in the figures mostly reflect uncertainties in the
$M/L$ ratio. We must not forget however, other error sources. For
example, Fig.~\ref{fig|MsLrMh}$b$ shows that different choices for
the GHMF yield a systematic difference in the results, that, at the
high-mass end, become comparable to the scatter considered in the
same figure. Further uncertainties come from estimates of the GSMF;
these are illustrated, in Figs.~\ref{fig|MsLrMh}, ~\ref{fig|Ratio},
~\ref{fig|Efficiency}, and ~\ref{fig|feedback}, by comparisons with
results obtained using the GSMF by Bell et al. (2003). Nevertheless,
our approach provides results consistent with observations, and have
comparable uncertainties. Moreover, since our approach bypasses any
assumption on the DM profile, it could provide a valuable tool to
discriminate among the different models of DM mass distribution in
galaxies.

Our analysis has shown that the relationships presented above bear
the imprint of the processes ruling the galaxy formation and
evolution. Models should eventually comply with them.

\begin{acknowledgements}
We thank S. Borgani, M. Girardi, and G.L. Granato for helpful
discussions, and the referee for a very careful reading of the
manuscript and many constructive comments that helped
substantially improving this paper. This work is supported by ASI
and MIUR grants.
\end{acknowledgements}

\clearpage

\appendix

\section{A simple feedback model}

The rate at which the gas mass $M_{\mathrm{inf}}$ falls toward the
central star forming regions can be written as
\begin{equation}\label{eq|dotMh}
\dot{M}_{\mathrm{inf}}(t)=-\frac{M_{\mathrm{inf}}(t)}{t_c}~.
\end{equation}
The infalling gas mass thus declines exponentially
$M_{\mathrm{inf}}(t) = M_{\mathrm{inf}}(0)\, \exp(-t/t_c)$, where
$t_c=\max[t_{\mathrm{cool}}(r_{\mathrm{vir}}),t_{\mathrm{dyn}}(r_{\mathrm{vir}})]$
is the maximum between the cooling and the dynamical time at the
virial radius, while $M_{\mathrm{inf}}(0)= f_{\mathrm{cosm}}\,M_h$
is the initial gas mass.

The time derivative of the cold, star forming gas mass is given by
\begin{equation}\label{eq|dotMc}
\dot{M}_{\mathrm{
cold}}(t)=\frac{M_{\mathrm{inf}}(t)}{t_c}-\psi(t)+R\psi(t)-\alpha
\psi(t)~,
\end{equation}
where $\psi(t)\equiv \dot{M}_{\mathrm{star}}$ is the Star
Formation Rate (SFR), $R$ is the fraction of mass restituted by
evolved stars ($R\approx 0.3$ for a Salpeter IMF), and
\begin{equation}\label{eq|SNfeed}
\alpha=\frac {N_{SN}\, \epsilon_{SN}\,E_{SN}}{E_B}
\end{equation}
is effective efficiency for the removal of cold gas by the
supernova feedback. Here $N_{SN}$ is the number of SNe per unit
solar mass of condensed stars, $\epsilon_{SN}\, E_{SN}$ the energy
per SN used to remove the cold gas, and $E_B$ the binding energy
of the gas within the DM halo, per unit gas mass. Following Zhao
et al. (2003) and Mo \& Mao (2004), the latter quantity can be
written as
\begin{equation}\label{eq|binding}
E_B=\frac {1}{2}\,V_{\mathrm{vir}}^{2}\,
f(c)(1+f_{\mathrm{cosm}})~,
\end{equation}
where $V_{\mathrm{vir}}$ is the circular velocity at the virial
radius for a halo mass $M_h$, $f(c)\approx 1$ is a weak function of
the concentration $c$, and we have assumed that, initially, the gas
fraction is equal to the cosmic baryon to dark-matter mass density
ratio, $f_{\mathrm{cosm}}$. Taking into account the dependence of
$V_{\mathrm{vir}}$ on the halo mass and redshift, we get, for $z\ga
1$:
\begin{equation}\label{eq|bindingvalue}
E_B\approx 3.2\times 10^{14} \left(\frac{1+z}{4}\right) \left
(\frac {M_h}{10^{12}\, M_{\odot}}\right)^{2/3} \,
(\mathrm{cm/s})^2~.
\end{equation}
The effective efficiency is well approximated by
\begin{equation}\label{eq|SNeff}
\alpha\approx 1.2 \, \left(\frac{N_{SN}}{8\times 10^{-3}}\right)\,
\left(\frac{\epsilon_{SN}}{0.1}\right)\,
\left(\frac{E_{SN}}{10^{51}\hbox{erg}}\right)\,  \left(
\frac{1+z}{4}\right)^{-1}\, \left(\frac{M_h}{10^{12}\,
M_{\odot}}\right)^{-2/3}~.
\end{equation}
Further setting
\begin{equation}\label{eq|SFR}
\psi(t)=\frac {M_{\mathrm{cold}}}{t_{\mathrm{star}}}~,
\end{equation}
$t_{\mathrm{star}}$ being the star-formation timescale,
eq.~(\ref{eq|dotMc}) is easily solved for $M_{\mathrm{cold}}(t)$.
The mass $M_{\mathrm{star}}$ cycled through stars is then
straightforwardly obtained, using eq.~(\ref{eq|SFR}):
\begin{equation}\label{eq|Ms}
M_{\mathrm{star}}(t)=\int_0^t\psi(t')dt'=\frac
{M_{\mathrm{inf}}(0)}{\gamma} \left[1- \frac{s\gamma}{s\gamma -1}
\, \exp(-t/t_c)+\frac {1}{s\gamma -1}\, \exp(-s\gamma
t/t_c)\right]~,
\end{equation}
where $\gamma=1-R+\alpha$. In the above formula,
$s=t_c/t_{\mathrm{star}} \gg 1$, since we expect that in the
central, clumpy regions the cooling and dynamical times are
shorter than $t_c$, which is estimated at the virial radius. The
mass in stars at the present time only includes the fraction
$f_{\mathrm{surv}}$ of stars still surviving:
$M_{\mathrm{star}}^{\mathrm{now}}=f_{\mathrm{surv}}\,
M_{\mathrm{star}}(t_{\mathrm{now}})$. The survived fraction
depends on the IMF and on the history of star formation; as a
reference, for a Salpeter IMF after about 10 Gyr from a burst we
have $f_{\mathrm{surv}}\approx 0.6$. If we assume that most of the
stellar feedback comes from SN explosions, then the total energy
injected into the gas is given by
\begin{equation}\label{eq|stellarfeed}
E_{\mathrm{star}}= \epsilon_{SN}\, E_{SN}\, N_{SN}\,
M_{\mathrm{star}}\approx 8\times 10^{58}\,
\left(\frac{\epsilon_{SN}}{0.1}\right)\,
\left(\frac{E_{SN}}{10^{51}\hbox{erg}}\right)\,
\left(\frac{N_{SN}}{8\times 10^{-3}}\right)\,
\frac{M_{\mathrm{star}}}{10^{11}\, M_{\odot}}~ \mathrm{erg}.
\end{equation}
As long as a significant amount of cool gas is present in the
central regions, we can imagine that there are mechanisms able to
remove angular momentum from gas clouds bringing them into a
reservoir around the central BH. One of these mechanisms is the
radiation drag (Kawakatu \& Umemura 2002) according to which, as
shown by Granato et al. (2004), the reservoir is fuelled at a rate
\begin{equation}\label{eq|dotMres}
\dot{M}_{\mathrm{res}}=1.2 \times 10^{-3}\psi(t)(1-e^{-\tau})~,
\end{equation}
where $\tau$ is the effective optical depth of the central star
forming regions [cf. Eqs.~(14), (15) and (16) of Granato et al.
2004]. If most of the mass in the reservoir is ultimately accreted
onto the central BH, we expect
\begin{equation}\label{eq|Mbh}
{M}_{BH}\approx 1.2 \times 10^{-3}\, M_{\mathrm{star}}\,
(1-e^{-\tau})~.
\end{equation}
Granato et al. (2004) assumed that the effective optical depth
depends on the cold gas metallicity and mass $\tau\propto Z\,
M_{\mathrm{gas}}^{1/3}$. The outcome of their numerical code
yields, on average, $Z\propto M_h^{0.3}$ in the mass range
$10^{11}\, M_{\odot}\leq M_h \leq 3 \times 10^{13}\, M_{\odot}$
[cf. their Figs.~(5) and (8)]. Since $M_{\mathrm{gas}}\sim
f_{\mathrm{cosm}} M_h$, one gets
\begin{equation}\label{eq|optdepth}
\tau \propto M_h^{2/3} ~.
\end{equation}
As for the AGN feedback, we use the prescription of Granato et al.
(2004); we rewrite their eq.~(28) for the kinetic luminosity that
can be extracted from AGN-driven winds, as
\begin{equation}\label{eq|wind}
L_K=\frac{1}{2}\, \dot{M}_w\,  v_{\infty}^2\approx 4.2 \times
10^{44}\, f_{c}N_{22}\, \left(\frac{M_{\rm BH}}{10^8
\,M_{\odot}}\right)^{1.5}~\hbox{erg}\,\hbox{s}^{-1},
\end{equation}
where $f_{c}$ is the covering factor of the wind and $N_{22}$ is the
gas column density in units of $10^{22}$ cm$^{-2}$. If we assume
that the BH mass is growing at around the Eddington rate, the total
kinetic energy in winds emitted by a BH of mass $M_{\rm BH}$ is
$E_{K}\approx (2/3)\, t_{ef}\, L_K(M_{\rm BH})$. This shows that the
action of the AGN occurs on a short timescale, the $e$-folding
timescale $t_{ef}=(\epsilon\, t_E)/(1-\epsilon)$ (where $t_E$ is the
Eddington time and $\epsilon$ is the BH mass to energy conversion
efficiency; for $\epsilon=0.1$, $t_{ef}\simeq 5\times 10^7\,$yr).

If a fraction $f_h$ of the AGN kinetic energy is transferred to
the gas, its total energy input is
\begin{equation}\label{eq|AGNfeed}
E_{\rm AGN}=f_h\, E_{K}\approx 3.6 \times 10^{60}\, f_h\,  f_{c}
N_{22}\, \left(\frac{\epsilon}{1-\epsilon}\right)\left({t_{E}\over
4\times 10^8\, \mathrm{yr}}\right)\, \left(\frac{M_{\rm BH}}{10^8
\, M_{\odot}}\right)^{1.5}\, \mathrm{erg}~.
\end{equation}
Since studies of BAL QSOs suggest that $N_{22}\geq 30$, $f_c\geq
0.1$ (see, e.g., Chartas et al. 2002; Chartas, Brandt \& Gallagher
2003) and $f_h\geq 0.3$ (see, e.g., Inoue \& Sasaki 2001; Nath \&
Roychowdhury 2002), we can take $f_h  f_c N_{22}\approx 1$. It is
interesting to compare this energy input to the total energy
released by accretion, $E_{\mathrm{acc}}=1.8 \times 10^{62}\,
(\epsilon/1-\epsilon)\, (M_{\rm BH}/10^8\, M_{\odot})$ erg:
\begin{equation}\label{eq|AGNfeedcomp}
E_{\rm AGN}\approx 2\times 10^{-2}\, E_{\mathrm{acc}}\,
\left(\frac{M_{\rm BH}}{10^8\, M_{\odot}}\right)^{0.5}~.
\end{equation}

\end{document}